# Impact of Tailored Gamma Irradiation on Pore Size and Particle Size of Poly [Ethylene Oxide] Films: Correlation with Molecular Weight Distribution and Microstructural Study

Madhumita Mukhopadhyay*[1], Mou Saha[1,2], Ruma Ray[2] and Sujata Tarafdar[1]

[1]Jadavpur University, Physics Department, Condensed Matter Physics Research Centre, Kolkata – 700032, WB, India

[2]Gurudas College, Physics Department, Kolkata – 700054, WB, India

Correspondence to: madhubanerji@gmail.com (M. Mukhopadhyay)
  **Tel:** +91 3324146666x2760, +91-9433882877
  **Fax:** +91 3324148917.

E-mail Addresses:  madhubanerji@gmail.com, sahamou14@gmail.com,
    me.rumaray@gmail.com and sujata_tarafdar@hotmail.com.

**Abstract**

The structure and morphology of polymers are significantly altered upon exposure to high energy gamma irradiation. The present investigation reports the influence of such irradiation of doses in the range of 1- 30 kGy on the particle size of Poly (Ethylene Oxide) [PEO] powder along with its correlation with molecular weight distribution. Pore-size distribution and overall porosity of the synthesized films prepared with irradiated and unirradiated PEO powder is also reported. The particle size of both unirradiated and irradiated PEO powders is found be in the range of $0.01\text{-}10^3$ μm. It is known that, variation in irradiation dose either generates particles of lower dimension through bond breakage i.e scission or higher dimension through cross-linkage. PEO films exhibit strong dependence of irradiation effect on pore size distribution and porosity. PEO films synthesized using both unirradiated and irradiated powders exhibit pore dimension in the range of 20 – 500 nm. The overall porosity of PEO films studied through BET adsorption technique initially increases with dose upto 3 kGy followed by linear decrease upto 30 kGy. Films cast with unirradiated PEO powder show multimodal pore size distribution, but perturbation with irradiation changes the modality to uni-or bimodal nature. The mentioned outcome of irradiation viz. particle and pore size variation is also correlated with PEO microstructures. The article demonstrates that, selective irradiation is capable of tailoring the pore-size within a definite regime thereby reducing the multimodal trait. The reported study may be relevant towards applications such as catalysis, sensing and filtration, where pore size distribution plays a crucial role.

# 1. Introduction

Polymer materials can be cast into films using many known techniques eg. solution casting, gel casting, emulsion freeze drying etc. The last two techniques are however limited to soluble or fusible polymers and cannot be conducted on some conjugate polymers. During the course of preparation and drying of such polymer films, pore phases are in-situ generated throughout the matrix even in absence of any pore former material [1]. In addition, numerous works have been cited on the preparation of porous polymer films [2-4]. Nano-or micro-porous polymer films are utilized in many applications eg. catalysis [5], sensing [6], filtration and fabrication of more complex nanostructured materials [7]. Nano-porous films attract the attention of present scenario due to certain advantages. These materials have low dielectric constants which are helpful in reducing the parasitic capacitance, signal crosstalk between interconnects and improve the switching speed in large scale integrated circuits [8]. Porosities of the films along with their distribution affect the associated dielectric, mechanical, thermal and chemical properties of the polymer and may enhance its feasibility for use in microelectronic technology. Enhanced porosity enables significant lowering of dielectric constant; however it degrades the mechanical and chemical properties of the developed film [9]. The magnitude of porosity should therefore be tailored so as to meet the requirement of the applied device along with the sustenance of other correlated properties. Presently, popular organic low dielectric constant polymer films without any permanent dipole moment allow lowering in permittivity value down to 2.5-2.8 [10]. Compared to this, porous dielectric inorganic or organic films could successfully reduce the permittivity magnitude below 2, thereby raising its popularity in application fields [10]. The porosity and its distribution could be determined by various techniques eg. spectroscopic ellipsometry and porosity/density simulation [for relative pore volume], mean pore size could be determined through X-ray reflectance, cross sectional focused ion beam combined with scanning electron microscopy technique, Rutherford backscattering spectroscopy and small-angle neutron scattering (SANS) [11-13]. In comparison to the aforementioned techniques, adsorption porosimetry is a simple and reliable newer technique for pore size distribution measurement [14]. BET (Brunauer–Emmet–Teller) adsorption technique involves different versions of porosimetry.

The present article intends to study the variation in porosity and pore size distribution of films prepared from Poly (Ethylene Oxide) powder subjected to high energy gamma irradiation of variable doses in the range of 1-30 kGy. Alterations in molecular level as an influence of irradiation are studied using molecular weight distribution pattern and correlated with other findings. The primary aim retains in the study of tailored irradiation effect within the lower dose range of 1-10 kGy. The variability in the dose of gamma irradiation in the powder polymer sample has been observed to significantly affect the pore size distributions of the developed films. Ionizing radiation is also reported to alter the microstructural pattern within the polymer matrix through the formation of new functional groups, rearrangements of bonds, scission and/or cross linking of chains etc. [15-18]. The major effects upon irradiation on bi-phasic polymer are scission or degradation producing chain fragments and cross linking or joining of chains by formation of bonds which occur simultaneously but usually one process dominates over the other depending on ambient conditions [19-22]. Influence of high energy irradiation, proton, electron beam, gamma rays, etc. on property alteration of polymer is reported by many researchers in prior arts [23-25]. Research is also presented so as to correlate the influence of variable molecular weight of polymer prepared using γ-radiation on its microstructure [26]. The impact of high energy irradiation on particle size distribution of polymer powder is also presented in the article along with its correlation with microscopic molecular weight distribution. Finally, microstructural studies of the films prepared with both unirradiated and irradiated PEO powder are reported.

## 2. Experimental Procedure

Pure powdered Poly [Ethylene Oxide] (PEO) from B.D.H., England (Mol Wt. $10^5$) was irradiated by gamma rays from a conventional gamma chamber using $^{60}$Co source with a dose rate 6.4 kGy.h$^{-1}$. The doses of gamma irradiation were varied in the range of 1-30 kGy in air. The irradiated powders were preserved in vacuum desiccator in air tight containers in order to protect from moisture and other external agents. Film preparation of experimental PEO was initiated with the preparation of sample solution in methanol. Two sets of samples with concentration of PEO as 2 % and 4 % (0.02 g.ml$^{-1}$ and 0.04 g.ml$^{-1}$) respectively were prepared. The solution was prepared by stirring the unirradiated and irradiated samples in the temperature range of 20-35$^o$C for ~ 16 h. These solutions of 2 wt. % and 4 wt. % concentrations were allowed to deposit in a polypropylene petri dish and dried in a desiccator to form thin films. The schematic representation for preparation of polymer films is given in Fig. 1. The figure also demonstrates the possible influence of gamma irradiation on polymer chains which results in simultaneous scission and cross linking phenomena.

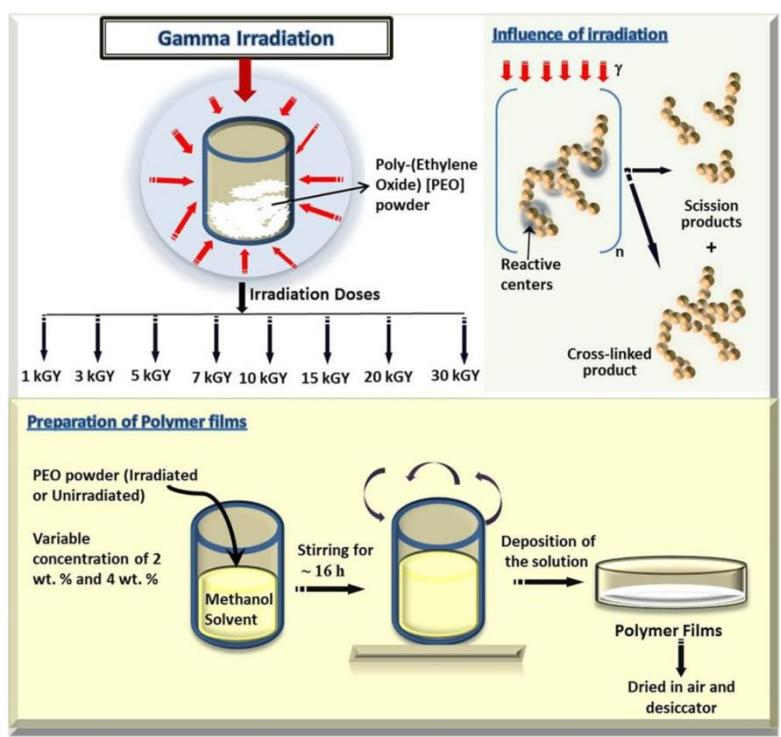

**Fig. 1.** Schematic representation for preparation of PEO films and influence of gamma irradiation on polymer chains.

The unirradiated and irradiated PEO powders were subjected to gel permeation chromatography (GPC) to study the molecular weight distribution, retention time and polydispersity index (PDI). For GPC, ~ 0.1 g.ml-1 of the experimental sample was dissolved in HPLC graded DMF (dimethyl formamide) using ultrasonic bath. Molecular weights and PDIs were measured by Waters gel permeation chromatography in DMF relative to poly methyl methacrylate (PMMA) and polystyrene (PS) standards on systems equipped with Waters Model 515 HPLC pump and Waters Model 2414 Refractive Index Detector at 3$^o$C with a flow rate of 1 mL.min-1. HRMS analyses were performed with Q-TOF YA263 high resolution (Waters Corporation) instruments by positive mode electrospray ionization.

The experimental powders were studied for particle size distribution using Mastersizer 2000 [Malvern instruments Pvt. Ltd., UK, Version 5.54, Serial Number: MAL1011591]. For such study, aqueous dispersion of ~ 1g of powder samples was poured into the Mastersizer inlet containing water vibrating at an adjustable rpm. The PEO films prepared from both unirradiated and irradiated powders were also subjected to BET analyzer [Quantachrome Instruments] for porosity and pore-size distribution. The above study is performed for PEO films of two different concentrations of 2 wt. % and 4 wt. %. Microstructural analysis was performed on the prepared films with the help of polarizable microscope [LAS EZ, Leica Application suit, Version: 2.0.0, 2010] to study the morphological effect of gamma irradiation. Image analysis of the obatined microstructure of the PEO films was performed using Image Pro software [Image Po Plus]. The reported experimental results are found to be reproducible for all the doses.

Experiments have also been performed upon subjecting the formed PEO films of variable concentration to gamma irradiation. The detailed study of the correlated properties of the irradiated PEO films will be communicated in due course.

## 3. Results and Discussion

### 3.1. Particle Size Distribution of Irradiated PEO Powder: Correlation with Molecular Weight Distribution

High energy gamma irradiation is found to result in simultaneous scission and cross linkage of the experimental polymer matrix. In the present context, gamma irradiation on polymer sample is subjected in air which being a scavenger promotes the scission compared to cross linkage. The predominance of scission could be clearly observed from Fig.2a which exhibits systematic reduction in the magnitude of both weight ($M_w$) and number ($M_n$) average molecular weight as a function of irradiation dose. $M_w$ and $M_n$ are derived from the molecular weight distribution (MWD) of unirradiated and irradiated PEO powders obtained from Gel Permeation Chromatography (GPC).

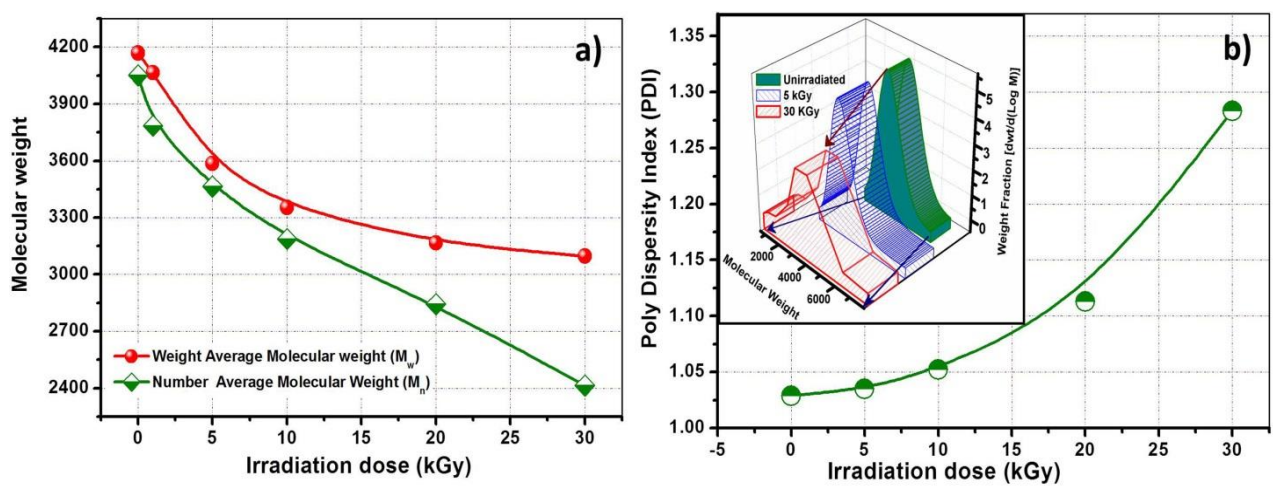

**Fig. 2.** a) Variation of weight average ($M_w$) and number average ($M_n$) molecular weight as function of irradiation dose and b) trend of poly dispersity index (PDI) obtained from GPC as a function of gamma irradiation dose; *Inset: Typical molecular weight distribution (MWD) pattern of unirradiated and irradiated (5 kGy and 30 kGy) PEO powders obtained from GPC.*

The representative MWD pattern of unirradiated and irradiated (5 kGy and 30 kGy) sample is shown in the inset of Fig. 2b. Detailed discussion on the influence of gamma irradiation on MWD pattern of PEO is discussed in our earlier communication [27]. Owing to the predominance of air-assisted scission phenomena, the associated cross linkage is not significant from the trend of $M_w$ and $M_n$ as a function of dose in Fig. 2a. Such involvement of cross linkage is visible from the MWD pattern (inset of Fig. 2b), in which small tail is formed in right side of the molecular weight in the irradiated PEO compared to the unirradiated sample (shown by an arrow at the right side of X-axis). With increase in irradiation dose, the peak of the GPC plot is observed to get lowered with maximum spreading towards low molecular weight and a slight fragment at the right. The progressive trend of poly dispersity index (PDI) as a function of irradiation dose (Fig. 2b) exhibits the heterogeneity of molecular fragments form as a result of higher scission contribution. However, the degree of heterogeneity is limited within 1.02-1.28, due to which the MWD pattern is significantly unimodal with slight bimodality generated at higher irradiation doses.

The particle size distribution of Poly [Ethylene Oxide] powder subjected to gamma irradiation of variable doses is shown in Fig. 3. The plots clearly represent the multimodal nature of particle size distribution within the polymer system unlike the MWD pattern (Fig. 2b inset). Hence, compared to microscopic MWD property, the macroscopic distribution of particle size is found to exhibit multidimensional polymer species as an effect of irradiation. The simultaneous contribution of scission and cross linkage with the predominance of the former as a consequence of irradiation perturbed PEO system is already established in above section (Fig. 2). From Figs. 3a-e, it can be said that irrespective of the magnitude of dose, the range of particle size remains constant; however, discrete doses generate particles of newer dimension (within the range 0.01 - $10^3$ μm). There is however no systematic change with irradiation dose. Small shifts in peak heights and position are seen but they appear erratic. The effect of irradiation on the distribution of particle sizes is summarized in Fig. 3f in which average peak area (A) is plotted as a function of irradiation dose. The approximate peak area

(A) is obtained from the product of change in particle size [dX ($X_2 - X_1$) where $X_2$ and $X_1$ are the particle size) at the half height of the peak with maximum volume % corresponding to the peak height. It can be clearly seen that, particle size distribution does not follow any definite pattern as a function of irradiation dose.

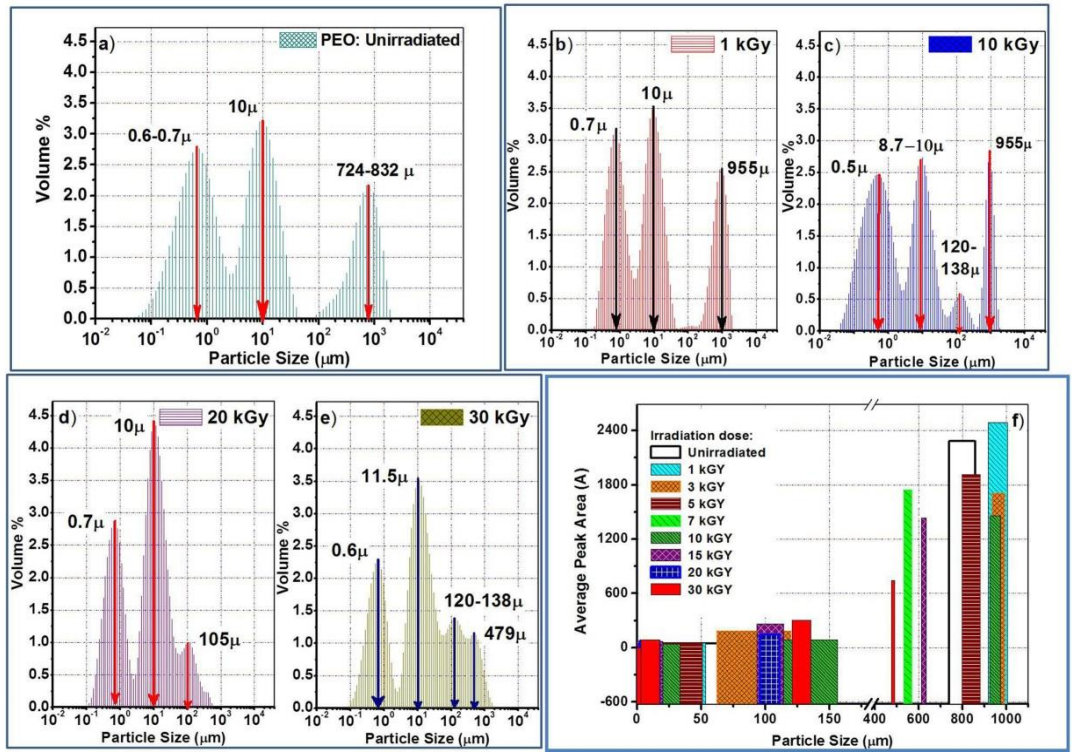

**Fig. 3.** Experimental particle size distribution plots for: a) unirradiated PEO; irradiated PEO with b) 1kGy, c) 10 kGy, d) 20 kGy, e) 30 kGy and f) variation of average peak area (A) as a function of particle size for variable irradiation doses.

Multimodal distribution of particle size is obtained with broad and overlapping peaks. The multimodality of the peaks of both unirradiated and irradiated PEO powders lies within definite range as mentioned above and varies erratically. It could be mentioned hereby that, range of particle size for high molecular weight PEO remains almost same with variation in its distribution as an effect of irradiation. Simultaneous role of scission and cross-linking with the predominance of the former during air assisted irradiation results for the variability in the distribution of particle size within constant regime of ~ 0.01 - $10^3$ μm.

## 3.2. A Study on the Influence of Irradiation on Porosity of PEO Film

The pore size distribution plots for PEO films prepared with both unirradiated and irradiated powder (dose: 1-30 kGy) having 2 wt. % and 4 wt. % concentrations is shown in Fig. 4 and Fig. 5. The figures represent pore volume distribution in which [dV/dDp] is plotted against pore size ($D_p$). The distribution function is such that area under the function in any pore diameter range is the volume (V) of pore in that range. The simultaneous occurrence of scission and cross linking as an influence of irradiation on PEO powder is already established through Gel permeation chromatography (GPC) in Fig. 2. GPC studies confirmed the predominance of scission phenomena during gamma irradiation (in air) through the formation of polymer species of lower molecular weight. Simultaneous contribution of cross-linkage is also supported from the presence of lower fraction of high molecular weight polymer from GPC. This is likely to affect the properties of prepared polymer films eg. the porosity variation studied through BET gas adsorption technique. PEO films synthesized using both unirradiated and irradiated powders exhibit pore dimension in the range of 20 – 500 nm. Comparing Figs. 4 and 5 with Fig. 3 shows that variable irradiation doses and unirradiated PEO sample exhibits the particle dimension in the range of $10^{-6}$ m, whereas, the formed pores in films are of much smaller dimension ($10^{-9}$ m). Both Fig. 4a and 5a exhibit multimodal distribution of pore size in case of films prepared with unirradiated PEO powder (2 & 4 wt. %). The extent of multimodality is higher in case of higher concentrated film of 4 wt. %. Irrespective of concentration variation, perturbation of such PEO powder with high energy gamma irradiation is found to reduce the number of modes in such multi-dimensional pore size distribution in films. In case of films with 2 wt. % PEO, the overall porosity is found to increase initially till 3 kGy followed by a steady reduction upto 30 kGy. In addition, with increase in irradiation dose, the pore size distribution tends towards unimodal behaviour with significant reduction in porosity.

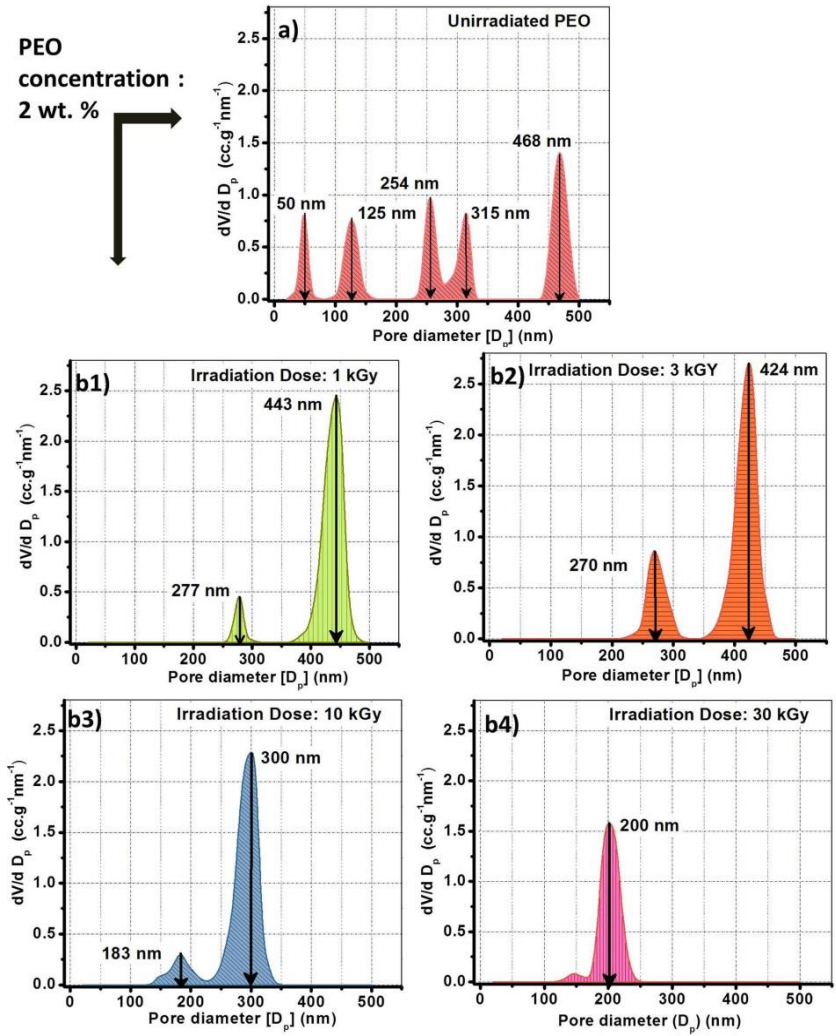

**Fig. 4.** Experimental pore size distribution plots for 2 wt. % films prepared using: a) unirradiated PEO; b) irradiated PEO with: b1) 1kGy, b2) 3 kGy, b3) 10 kGy and b4) 30 kGy.

Increase in irradiation also tend to shift the pore diameter towards smaller dimension i.e shifting of the plots from right to left side with increasing radiation from 1 kGy to 30 kGy. Similar trend in pore-size distribution is observed in case of higher concentration of 4 wt. % as shown in Fig. 5. This could be explained on the fact that, increment in irradiation generates increased number of radicals or excited species within the nuclei and globules [28]. The increase in number of excited components stabilizes through diffusion of polymer chains towards the existing nuclei forming larger spherullites. Consequently, smaller voids are obtained within large formed sperullites resulting in shifting of pore-size distribution towards lower dimension. Compared to Fig. 4, irradiated films with higher PEO concentration (4 wt. %) are found to exhibit tri-modal pore distribution below 10 kGy which changes to bi-modal till highest dose of 30 kGy. In addition, the polymer films with 4 wt. % concentration are found to be less porous compared to 2 wt. %. This is shown in Fig. 6 which exhibits a comparison of % porosity as a function of irradiation dose for the films prepared with 2 wt. % and 4 wt. % PEO solution.

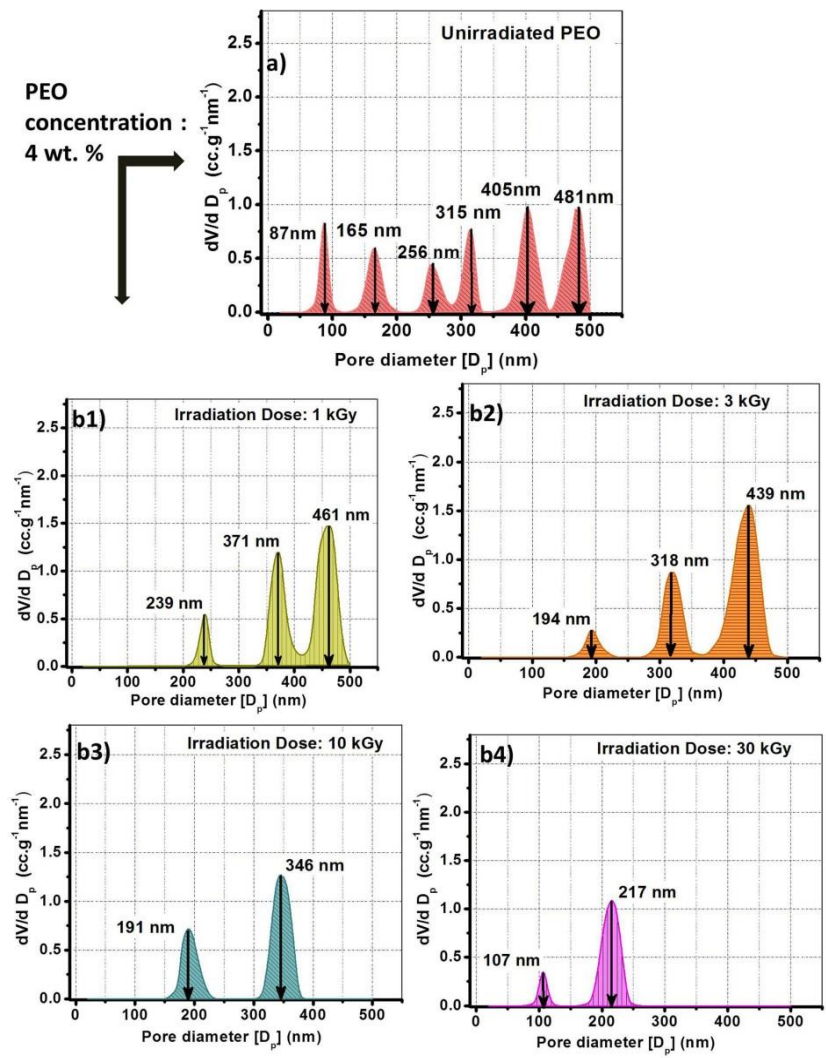

**Fig. 5.** Experimental pore size distribution plots for 4 wt. % films prepared using: a) unirradiated PEO; b) irradiated PEO with: b1) 1kGy, b2) 3 kGy, b3) 10 kGy and b4) 30 kGy.

Fig. 6 exhibits that, irradiation dose upto 3 kGy increase the porosity of the matrix to 34 % and 25 % for 2 wt. % and 4 wt. % PEO concentrations respectively. This is followed by a linear decrease in film porosity upto 30 kGy [10 % for 2 wt. % and 6 % for 4 wt. %]. The irradiation being carried out in air, higher doses tend to favor scission predominance leading to densification of polymer matrix thereby lowering the overall porosity. The resultant porosity of the films prepared with irradiated powder of higher doses is therefore found to be smaller and pore sizes are in the range of 100 nm. It could be stated that at lower concentration (2 wt. %), due to the presence of lesser polymer matrix per unit volume, only intra-molecular cross linking is feasible in solution subjected to constant stirring [27]. However, in higher concentration of 4 wt. %, both intra-and inter-molecular cross linkage is feasible, which tend to reduce the overall porosity of the matrix significantly as evident from Figs. 4-6. However, the irradiation being carried out in air, contribution of simultaneous scission is significant which generates particles of lower dimension and thereby generates pores within two regimes in Fig. 4 and three regions in Fig. 5.

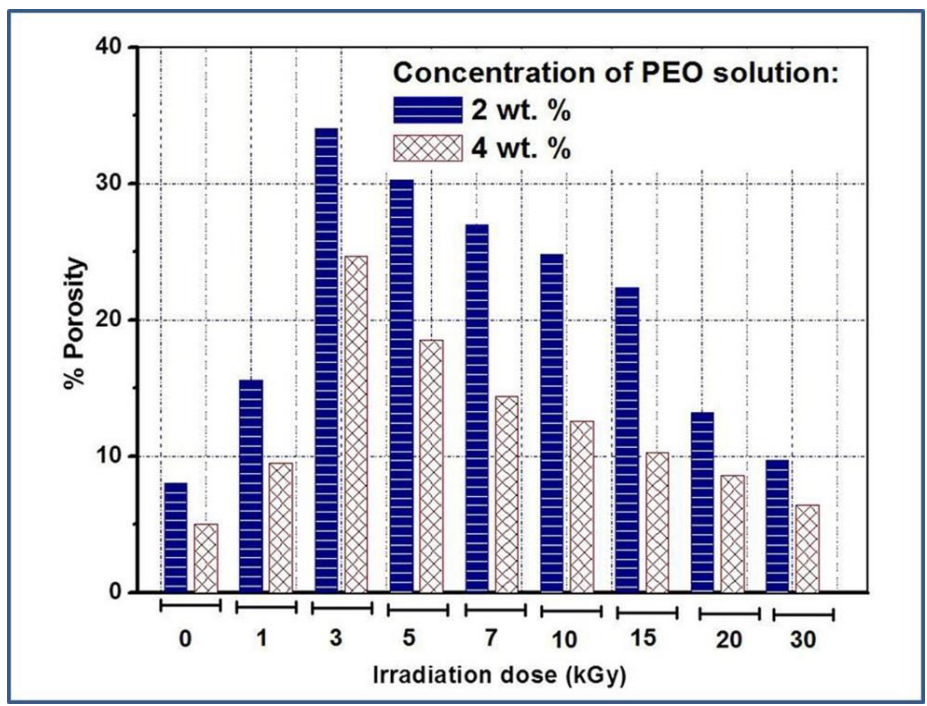

**Fig. 6.** Comparison of % porosity for films prepared with variable PEO concentration of 2 wt. % and 4 wt. % as a function of irradiation dose.

Fig. 7 summarizes the data in Fig. 4 to facilitate comparison among pore dimension of films prepared with irradiated and unirradiated PEO powders of 2 wt. % concentration. This graph is plotted considering the pore volume distribution ($dV/dD_p$) of corresponding pore distributions at the peak maxima. Films prepared with unirradiated powders shows multimodal pore size distribution (represented by white bars in Fig. 7). It could be observed that, two regimes (stronger peak and weaker peak) exist for irradiation dose of 1-20 kGy.

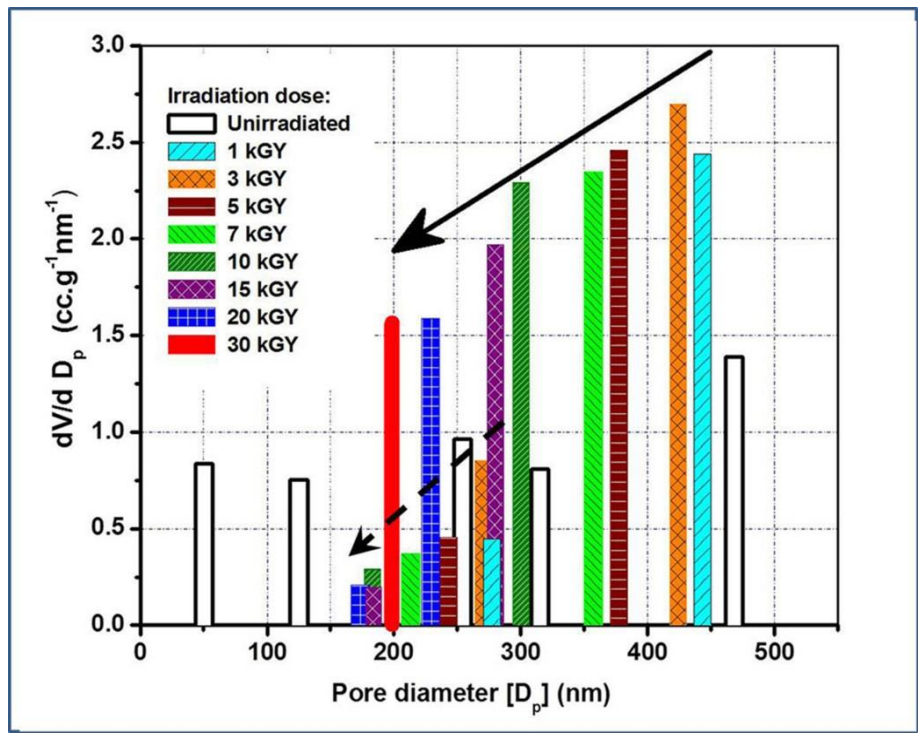

**Fig. 7.** Comparative pore dimensions of films prepared with unirradiated and irradiated PEO powder of 2 wt. % concentration.

*The white bars represent the multimodal pore size distribution in unirradiated PEO. Irradiation leads to a bimodal distribution with stronger peak at higher pore dimension and a weaker peak at a lower pore size. On irradiation, the higher sized peak and the lower sized peak shift towards lower pore dimension systematically with dose. This is shown by solid (for stronger peak) and dashed (for weaker peak) arrows respectively.*

Two peaks [stronger and weaker] of the bimodal distribution for each irradiation dose are shown. Shifting of the pore sizes corresponding to both stronger (represented by solid arrow) and weaker (represented by dashed arrow) from right to left shows systematic change with increase in dose from 1 to 20 kGy.

However, *particle size distribution* does not follow any definite trend as a function of irradiation doses unlike distribution of pore size. The ordered trend observed in Fig. 7 is absent in the particle size distribution as shown in Fig. 3f.

**3.3. Microstructural Study**

In polymer physics, spherullites are spherical semi crystalline regions inside non-branched linear polymers. Their formation is associated with crystallization of polymers from the melt and is controlled by several parameters such as the number of nucleation sites, structure of the polymer molecules, cooling rate, etc. Spherulites are composed of highly ordered lamellae, which are connected by amorphous regions which provide certain elasticity and impact resistance. Alignment of the polymer molecules within the lamellae results in birefringence producing a variety of colored patterns, including Maltese cross, when spherullites are viewed between crossed polarizers in an optical microscope.

Microstructures of unirradiated and irradiated PEO films along with their software analysis images are given in Figs. 8-10. The sperullites formed in films prepared with unirradiated powder are not well defined and are arranged in disordered manner (Fig.8a). The matrix of unirradiated polymer is found to be highly amorphous i.e unstructured with abrupt formation of spherullite at non-deterministic positions without any regularity. This feature is more clarified from the binerized image of Fig. 8b and is further supported from the irregular distribution of intensity at selected locations (Fig. 8a1). This is in parity with the particle size distribution of unirradiated PEO (Fig. 3a) which elucidates the presence of multimodal particle dimensions of unified volume percentage. This signifies that in unirradiated

/unperturbed polymer medium, the particle fails to form a structured geometry and the matrix is more amorphous.

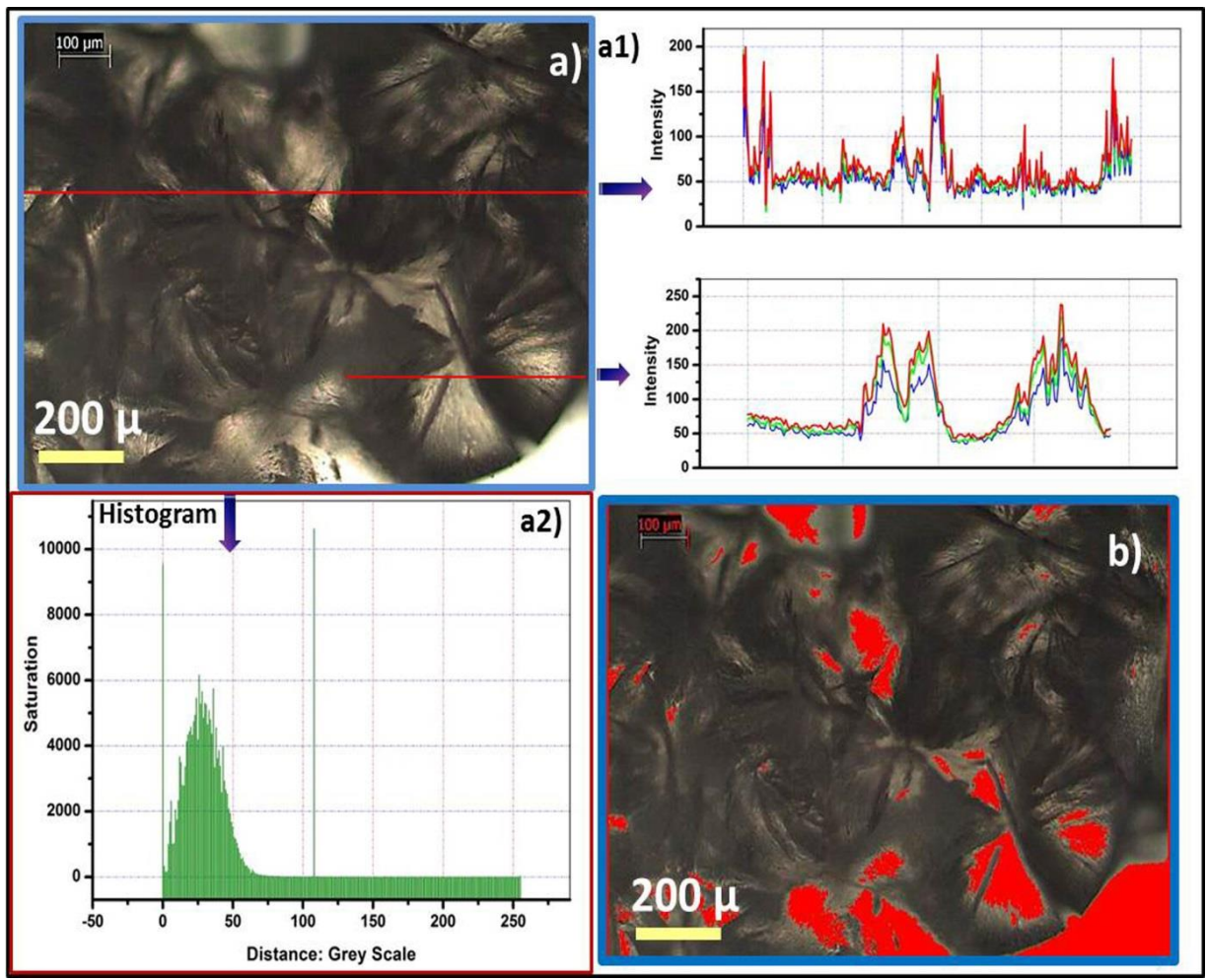

**Fig. 8.** a) Microstructure of unirradiated PEO film along with it's: a1) line profile intensity distribution, a2) saturation vs. grey scale histogram and b) binerized image generated from bright objects counting.

However, polymer perturbed with gamma irradiation undergoes simultaneous scission and both intra-and inter-particle cross linkage which helps in the generation of spherullite of finite structure and allows its distribution throughout the polymer regime. The dimension and linkage of these formed spherullite are nevertheless, dependent on irradiation dose.

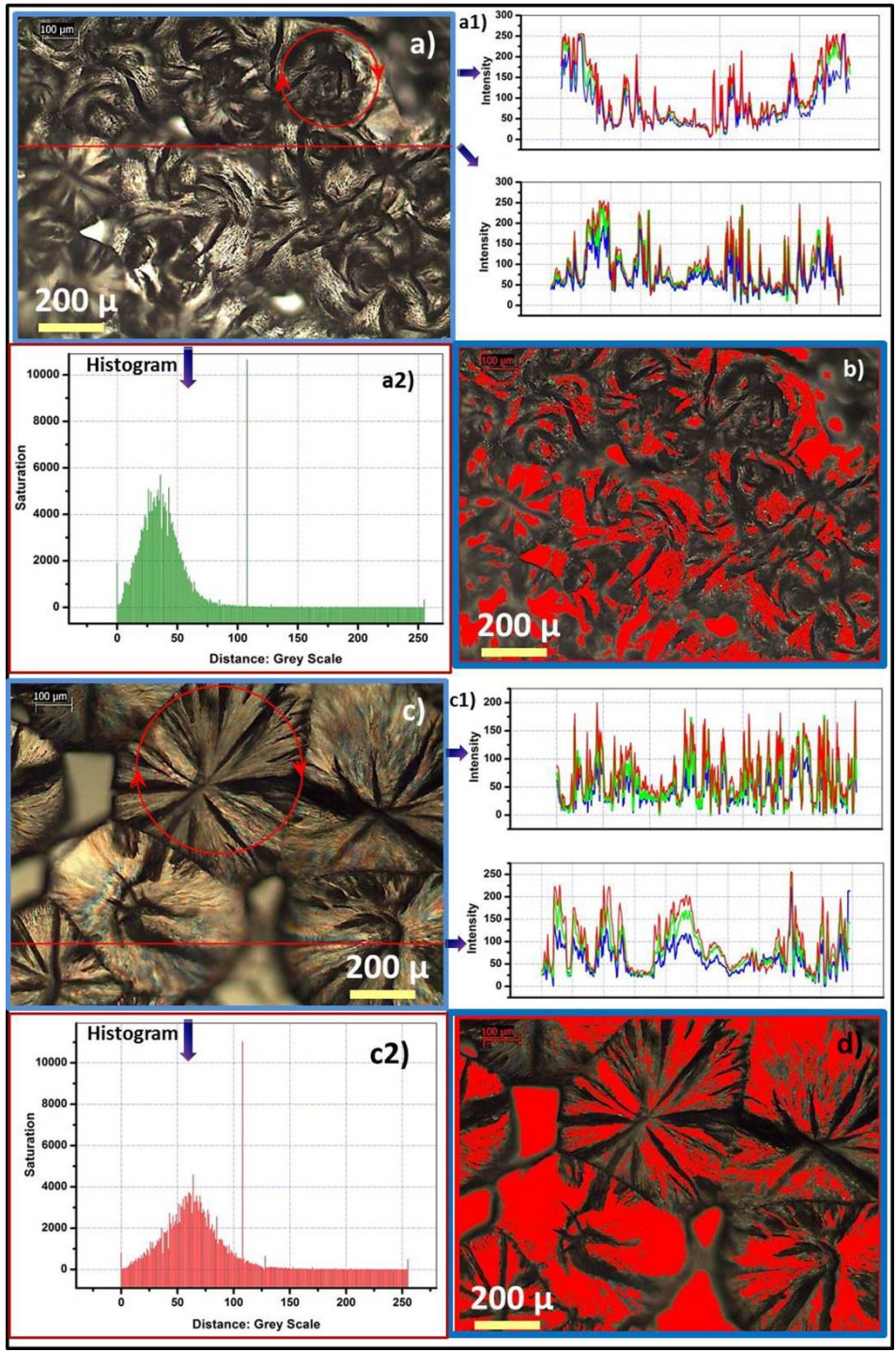

**Fig. 9.** Analysis of 2 wt. % PEO film with: a) microstructure of the irradiated film at 5 kGy along with its: a1) line profile intensity distribution, a2) saturation vs. grey scale histogram and b) binerized image; c) microstructure of the irradiated film at 20 kGy along with its: c1) line profile intensity distribution, c2) saturation vs. grey scale histogram and d) binerized image.

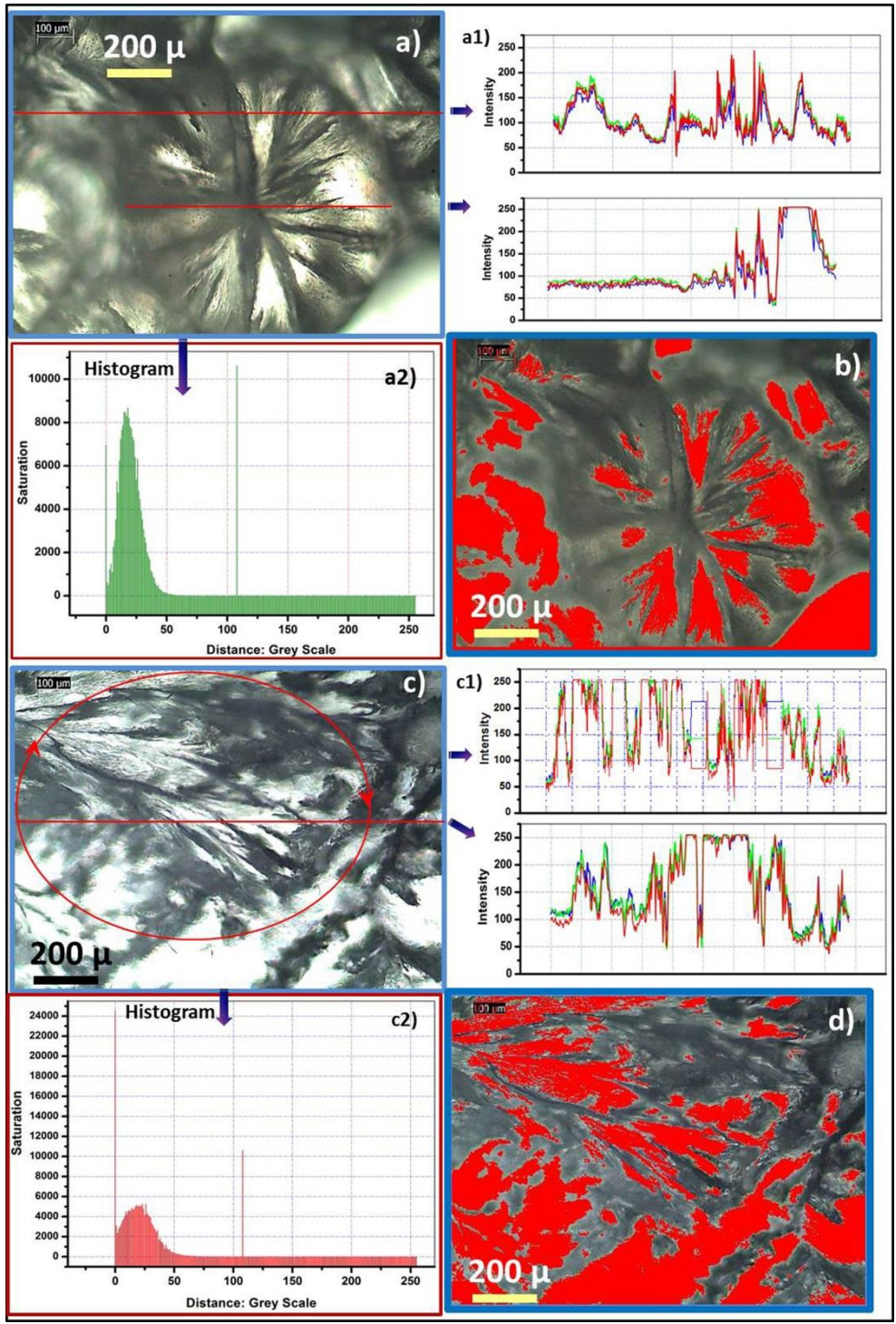

**Fig.10.** Analysis of 4 wt. % PEO film with: a) microstructure of the irradiated film at 5 kGy along with its: a1) line profile intensity distribution, a2) saturation vs. grey scale histogram and b) binerized image; c) microstructure of the irradiated film at 20 kGy along with its: c1) line profile intensity distribution, c2) saturation vs. grey scale histogram and d) binerized image.

Microstructures of unirradiated and irradiated PEO films along with the variation of spherullite dimension as a function of irradiation dose are given in Fig. 11.

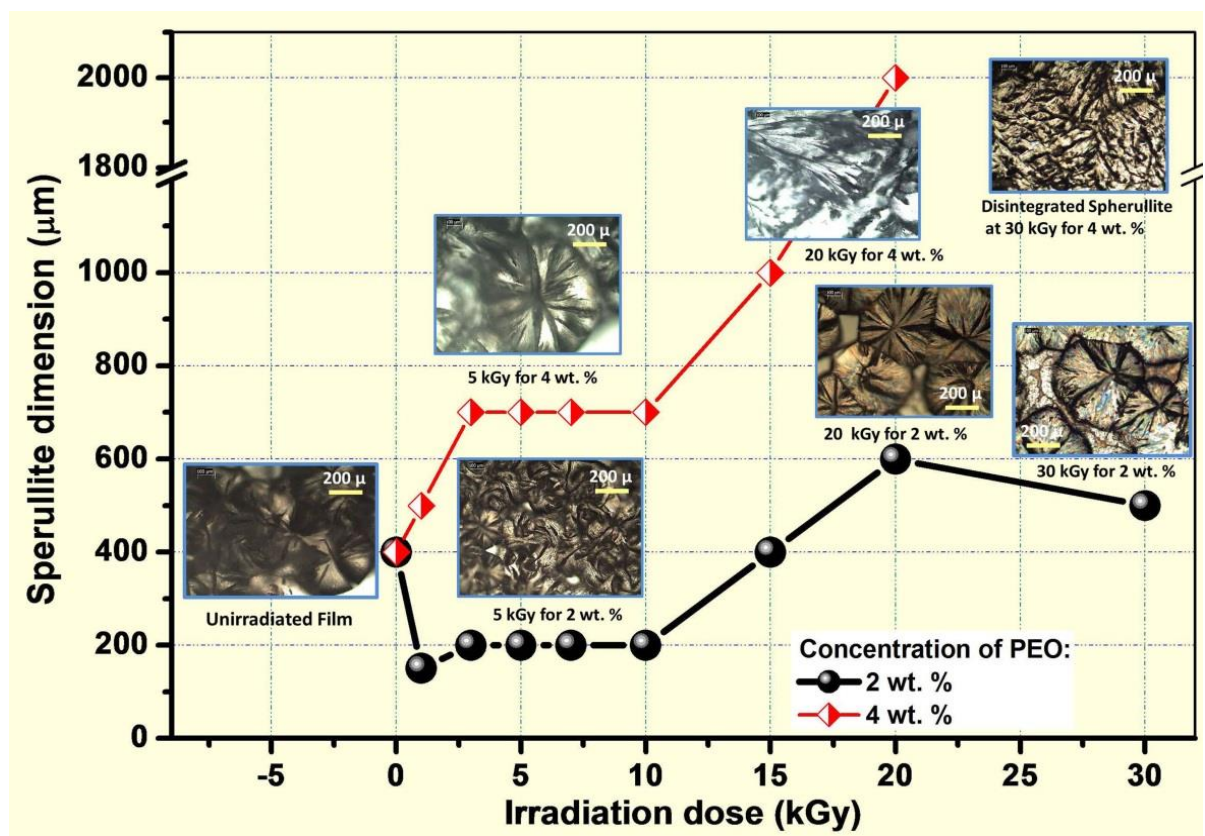

**Fig. 11.** Variation in spherullite dimension in for microstructures of PEO film as a function of irradiation dose. The inset pictures represent the microstructures of films prepared with irradiated PEO powder.

It could be observed from Fig. 9 a and c, that irrespective of similar concentration of 2 wt. %, the microstructural pattern varied significantly with increasing irradiation dose from 5 kGy to 20 kGy. At higher irradiation doses, distinct spherical spherullite of dimension ~ 400-500 μm is formed (Fig. 11). At 5 kGy dose, such spherical entities of lower size (≤100 μm) are visible; however, their distribution pattern is non-uniform throughout the matrix. It could be more precisely observed from the binerized images (Fig. 9 b and d), that, at lower radiation dose, the extent of cross linkage prior to scission of the fragments is less, owing to which formation of spherical polymer entities is incomplete and are of significantly lower dimension. In contrast, at higher doses, contribution of cross linkage is higher along with scission dominance and thereby generates polymer spherullite of higher dimension. Though the polymer system under investigation is a unified PEO system, but the intensity distribution of selected zones (Fig. 9 a1 and 9 c1) elucidates the presence of two distinct regions within the matrix. Unified distribution of intensity is observed at the selected regions of 2 wt. % PEO irradiated at 20 kGy (Fig. 9a1) compared to Fig. 9c1, which signifies the formation of uniform polymer spherullite through the matrix, whereas, at lower irradiation dose, the formation of these spherullite seems to have non-uniform distribution. The quantitative image analysis of saturation intensity as a function of distance shows much wider distribution with reduced saturation peak intensity for PEO samples at higher irradiation (Fig. 9c2). In addition, the peak of the distribution is found to get shifted at higher distance with increase in radiation dose. The influence of concentration of perturbed polymer system could be observed from Fig. 10, which shows the microstructure and image analysis of 4 wt. % polymer film. Unlike in 2 wt. % concentration, the dimension of spherical spherullite is observed to increase significantly at either 5 kGy (Fig. 10 a) or 20 kGy (Fig. 10 c) at 4 wt. % polymer concentration. This could be supported from the intra-molecular linkage possible at lower concentration and both intra-and inter-molecular linkage at higher concentration of 4 % as illustrated in our earlier article [27]. Unlike in 2 wt. % concentrations, the dimension of spherical spherullite is observed to increase significantly for films of 4 wt. % concentrations at all irradiation doses (500 µm to 2000 µm for 1 kGy to 20 kGy). The polymer chains in 4 wt. % concentration are no longer independent but get entangled to make the situation complex. The extent of increment in spherullite dimension is found to increase with irradiation dose till 20 kGy (~ 2000 µm). At 20 kGy dose, spherical spherullite of much higher dimension is formed and a part of it's could only be visible under the present frame. The dimension of the spherullite corresponding to 20 kGy dose (4 wt. %) reported in Fig. 11 is an approximate reconstruction. Further increase in dose to 30 kGy reduces the spherullite dimension for 2 wt. %. In contrast, such a high dose is found to disintegrate the formed sperullites at 4 wt. % concentration, owing to which the dimension is not reported in Fig. 11. Since, the zones for line profile are selected within a part of one enlarged spherullite for 20 kGy (4wt. %), the distribution of intensity are rather inconsistent (Fig. 10 c1). The binerized images at 5 kGy and 20 kGy dose for 4 wt. % film (Fig. 10 b, d) illustrates the formation of much higher dimension spherullite with individual distribution of bright and dark phases. Alike, Fig. 9 a2 and c2, the distribution of saturation in histogram for higher concentrated PEO film follows similar trend at 5 kGy and 20 kGy in Fig. a2 and c2. At 5 kGy, the distribution shows much sharp peak with enhanced saturation, however, the distribution is observed to reduce significantly with slight shifting at higher distance for higher irradiation dose of 20 kGy.

4. Conclusion

The present article is focused on the influence of high energy gamma irradiation in terms of tailored doses on the property evaluation of the poly [Ethylene Oxide] (PEO) films synthesized using such irradiated powders. PEO powder is exposed to selective doses of gamma irradiation within 1-30 kGy, tailored within lower regime of 1-10 kGy [i.e. 1kGy, 3kGy, 5kGy and 10kGy]. The predominance of scission in air assisted gamma irradiation is clarified from the systematic reduction in the magnitude of both weight ($M_w$) and number ($M_n$) average molecular weight obtained from Gel Permeation Chromatography (GPC) as a function of irradiation dose. The simultaenous involvement of cross linkage is visible from the MWD pattern, in which small tail is formed in right side of the molecular weight in the irradiated PEO compared to the unirradiated sample. Particle size distribution of all the experimental unirradiated and irradiated PEO samples exhibits a multimodal distribution pattern. Variable irradiation doses are found to produce either smaller or large polymer particles; however, the dimensional range of particles remains constant within 0.1 to $10^3$ μm.

Multimodal peaks are obatined which are broad and overlapping in character. We conclude that, the influence of irradiation on particle size distribution of PEO powder is not very systematic. In contrast to the dimension of particles ($10^{-6}$ m), pore sizes of the films are found to be in the nanometric ($10^{-9}$ m) range. It is likely that particle size of the experimental polymer powder are not directly correlated with pore size of the polymer films synthesized using after dissolving such powders in a solvent. Study of pore – size distribution is conducted on polymer films prepared using both unirradiated and irradiated PEO powder using varying concentration of 2 wt. % and 4 wt. %. The films prepared with 4 wt. % PEO are found to be less porous compared to lower concentration of 2 wt. %. Polymer films with unirradiated powders exhibit multimodality in pore size which tends to convert towards unimodal for 2 wt. % and bi- or tri-modal for 4 wt. % concentrations. Increase in irradiation also tends to shift the pore diameter towards smaller dimension. Unlike particle size, the pore size distribution of the films exhibit significant and systematic variation with irradiation dose. The films prepared with either 2 wt. % or 4 wt. % shows an initial increase in the overall porosity upto 3 kGy [34 % for 2 wt. % and 25 % for 4 wt. % concentration] which subsequently decreases till the highest dose of 30 kGy [10 % for 2 wt. % and 6 % for 4 wt. %]. The films prepared with irradiated powder of higher doses tend to densify the matrix and generates smaller pores in the range of 100 nm for both the concentration variations.

The aforementioned outcomes are correlated with microstructural study of the developed films. Polymer films with unirradiated PEO exhibits disordered spherullite growth where the boundaries are not well defined. In contrast, gamma irradiation generates distinct spherullite and significant correlation among its size with irradiation dose. Irrespective of the concentration variation, dimension of the sperullites follows an increasing trend with irradiation dose upto 20 kGy. Further increase in irradiation dose to 30 kGy results in reduced spherullite dimension for 2 wt. %. However, irradiation dose of 30 kGy disintegrates the formed spherullite in films prepared with 4 wt. % concentration. In summary, selective irradiation is capable of tailoring the pore-size within a definite regime thereby reducing the multimodal trait. The reported study may be relevant towards applications such as catalysis, sensing and filtration, where pore size distribution plays a crucial role. Further study on the complete understanding of the correlation between the mentioned observations is underway.

**Acknowledgements**

The research work is funded by UGC Major Research Project [F. No. 41-847/2012 (SR)] for financial support and MM is thankful to Council of Scientific and Industrial Research (CSIR), India for providing Research Associate fellowship. Dr. Paramita Bhattacharyya and Miss. Suchisrawa Ghosh of Department of Food Technology, Jadavpur University, Kolkata are acknowledged in extending their support in using the gamma irradiation chamber. Dr. T. K. Pal and Mr. Subhashis Dan of Bioequivalence study Centre, Department of Pharmaceutical Technology, Jadavpur University, Kolkata are acknowledged for their co-operation in using the Particle size analyzer. Dr. Raja Shunmugam and Mr. Vijay Rao of Indian Institute for Science Education & Research (IISER), Kalyani, India are also acknowledged for their help regarding molecular distribution study by GPC technique.

**References**

1. Liang Li, F. Yan, Gi Xue. J. Appl. Polym. Sci. **91**, 303- 307(2004).

2. G. Liu, Jianfu Ding. Adv. Mater. **10**, 69-71 (1998).

3. M. Yoshida, M. Asano, A. Safranj, H. Omichi, R. Spohr, J. Vetter, R. Katakai, Macromolecules. **29,** 8987-8989 (1996).

4. R. A. Zoppi, S. Contant, E. A. R. Duek, F. R. Marques, M. L. F. Wada, S. P. Nunes, Polymer. **40**, 3275-3289 (1999).

5. H. P. Kormann, G.Schmid, K. Pelzer, K. Philippot, B. Chaudret, Z. Anorg. Allg. Chem. **630**, 1913-1918 (2004).

6. S. Pan, L. J. Rothberg, Nano Lett. **3**, 811-814 (2003).

7. K. Fukutani, K. Tanji, T.Motoi, T. Den, Adv. Mater. **16**, 1456 -1460 (2004).

8. N. Aoi, Jpn. J. Appl. Phys. **36,** 1355 -1359 (1997).

9. Z. Y. Yang, M. Zhao. J. Opt. A-Pure Appl. Op. **9**, 872-876 (2007).

10. M. R. Baklanova, K. P. Mogilnikov, V. G. Polovinkin, F. N. Dultsev, , J. Vac. Sci. Technol., B. **18**, 1385-1391 (2000).

11. K. P. Mogilnikov, V. G. Polovinkin, F. N. Dultsev, M. R. Baklanov, MRS Proceedings, 565-581 (1999). DOI:10.1557/PROC-565-81.

12. S. Dourdain, J. F. Bardeau, M. Colas, B. Smarsly, A. Mehdi, B. M. Ocko, A. Gibauda, Appl. Phys. Lett. **86**, 113108-1--113108-3 (2005).

13. J. H. Yim, Y. Y. Lyu, H. D. Jeong, S. A. Song, I. S. Hwang, J. H. Lee, S. K. Mah, S. Chang, J. G. Park, Y. F. Hu, J. N. Sun, D. Gidley, Adv. Funct. Mater. **13**, 382-386 (2003).

14. S. J. Gregg, S. W. Sing, in Adsorption, Surface Area and Porosity, 2nd Edn. (Academic, New York, 1982).

15. Zainuddin, J. Albinska, P. Ulanski, J. M. Rosiak, J. Radioanal. Nucl. Chem. **253**, 339-344 (2002).

16. U. Gryczk, W. Migdal, D. Chmielewska, M. Antoniak, W. Kaszuwara, A. Jastrzebska, A. Olszyna, Radiat. Phys. Chem. **94,** 226- 230(2014).

17. A. A. Abiona, A. G. Osinkolu, Int. J. Phys. Sci. **5,** 960-967 (2010).

18. E. M. Kempner, J. Polym. Sci., Part B: Polym. Phys. **49**, 827-831 (2011).

19. A. Charlesby, in Atomic Radiation and Polymers. (Pergamon Press: Oxford, UK. 1960).

20. M. Dole, (Ed.) in The Radiation Chemistry of Macromolecules. (Academic Press: New York, Vol.1, 1972).

21. S. Ghosal, R. Ray, T. K. Ballav, S. Tarafdar, Indian J. Pure Appl. Math., **51**, 324-332 (2013).

22. S. Ghosal, M. Mukhopadhyay, R. Ray, S. Tarafdar, Physica A. **400**, 139-150 (2014).

23. K. Schrote, M. W. Frey, Polymer. **54,** 737-742 (2013).

24. B. Croonenborghs, M. A. Smith, P. Strain, Radiat. Phys. Chem. **76,** 1676- 1678(2007).

25. P. Nanda, S. K. De, S. Manna, U. De, S. Tarafdar, Nucl. Instrum. Methods Phys. Res., Sect. B. **268**, 73-78 (2010).

26. Joana J.H. Lancastre, António N. Falcão, Fernanda M.A. Margaça, Luís M. Ferreira, Isabel M. Miranda Salvado, Maria H. Casimiro, László Almásy, Anikó Meiszterics, Radiat. Phys. Chem. **106,** 126- 129(2015).

27. M. Saha, M. Mukhopadhyay, R. Ray, T. K. Ballabh, S. Tarafdar. Under Revision in Modell. Simul. Mater. Sci. Eng. 2014.

28. P. S. Clair, Macromolecules. **42,** 3469-3482 (2009).